\documentclass[fleqn,10pt]{wlscirep}

\title{Unexpected 3+ valence of iron in  FeO$_2$, a geologically important
material lying ``in between'' oxides and peroxides}

\author[1,2,*]{Sergey S. Streltsov}
\author[1,2]{Alexey O. Shorikov}
\author[1,2]{Sergey L. Skornyakov}
\author[1]{Alexander I. Poteryaev}
\author[3]{Daniel I. Khomskii}
\affil[1]{M.N. Miheev Institute of Metal Physics of Ural Branch of Russian Academy of Sciences, Ekaterinburg, Russia}
\affil[2]{Ural Federal University named after the first President of Russia B.N.Yeltsin, Theoretical Physics and Applied Mathematics Department, Ekaterinburg, Russia}
\affil[3]{II. Physikalisches Institut, Universit\"at zu K\"oln, K\"oln, Germany}

\affil[*]{streltsov@imp.uran.ru}

\begin{abstract}
Recent discovery of FeO$_2$, which can be an important
ingredient of the Earth's lower mantle and which in particular may
serve as an extra source of water in the Earth's interior, opens new perspectives for geophysics and geochemistry, but this is also an extremely interesting material from physical point of view. We found that in contrast to naive expectations Fe is nearly 3+  in this material, which strongly affects its magnetic properties and makes it qualitatively different from well known sulfide analogue - FeS$_2$. Doping, which is most likely to occur in the Earth's mantle, makes FeO$_2$ much more magnetic. In
addition we show that unique electronic structure places FeO$_2$ ``in
between'' the usual dioxides and peroxides making this system
interesting both for physics and solid state chemistry.
\end{abstract}
\begin{document}

\flushbottom
\maketitle

\thispagestyle{empty}

Recent discovery of a new iron oxide FeO$_2$, which does not exist at normal conditions, but can be stabilized at a very high pressure (76 GPa) and temperature (1800 K)\cite{Hu2016} may dramatically shift our understanding of how Earth is formed and what was a source of water in interior of our planet. FeO$_2$ is expected to appear in the Earth's lower mantle below 1800 km\cite{Hu2016} and start to dominate over other Fe oxides at higher pressures. The composition of the mantle is extremely important for the seismology, since it determines convection processes. There were proposed a number of structural models based on different ratio of ferropericlase (a solid solution of FeO and MgO), bridgmanite (Mg,Fe,Al)(Al,Fe,Si)O$_3$, (Mg,Fe)$_2$SiO$_4$ olivine and other compounds\cite{DZIEWONSKI1981297,Irifune2010,Murakami2012,Wang2015}, but none of them took into account existence of FeO$_2$. Moreover, physical properties of this material are completely unexplored.  One might expect that they can be highly unusual, since on  one hand Fe ion in FeO$_2$ formally should have exceptionally high oxidation state, 4+. Since the O-O distance in FeO$_2$ is 1.89\AA~it is not likely that there can be a strong bonding between the O ions, like in molecular oxygen (where the O-O bond distance is 1.21\AA) and one may indeed expect that Fe ions will adopt 4+ valence state and then FeO$_2$ is in a negative charge transfer regime\cite{Khomskii-97,Sawatzky2016,khomskii2014transition,ZSA}. This may result in self-doping \cite{Korotin1998} and also to  bond or charge disproportionation\cite{Bisogni2016,Kawakami2016}, inversion of the crystal field splitting\cite{Ushakov2011} or nontrivial magnetic structures. On the other hand, the presence of the ligand-ligand dimers may also strongly affect physical properties of FeO$_2$ as it does in the  actual pyrite FeS$_2$  (``the fool's gold''). However, O-O distance in FeO$_2$ is 1.89\AA, much larger than in molecular oxygen (1.21\AA).

\begin{figure}[t!] 
 \centering
\includegraphics[width=1\columnwidth]{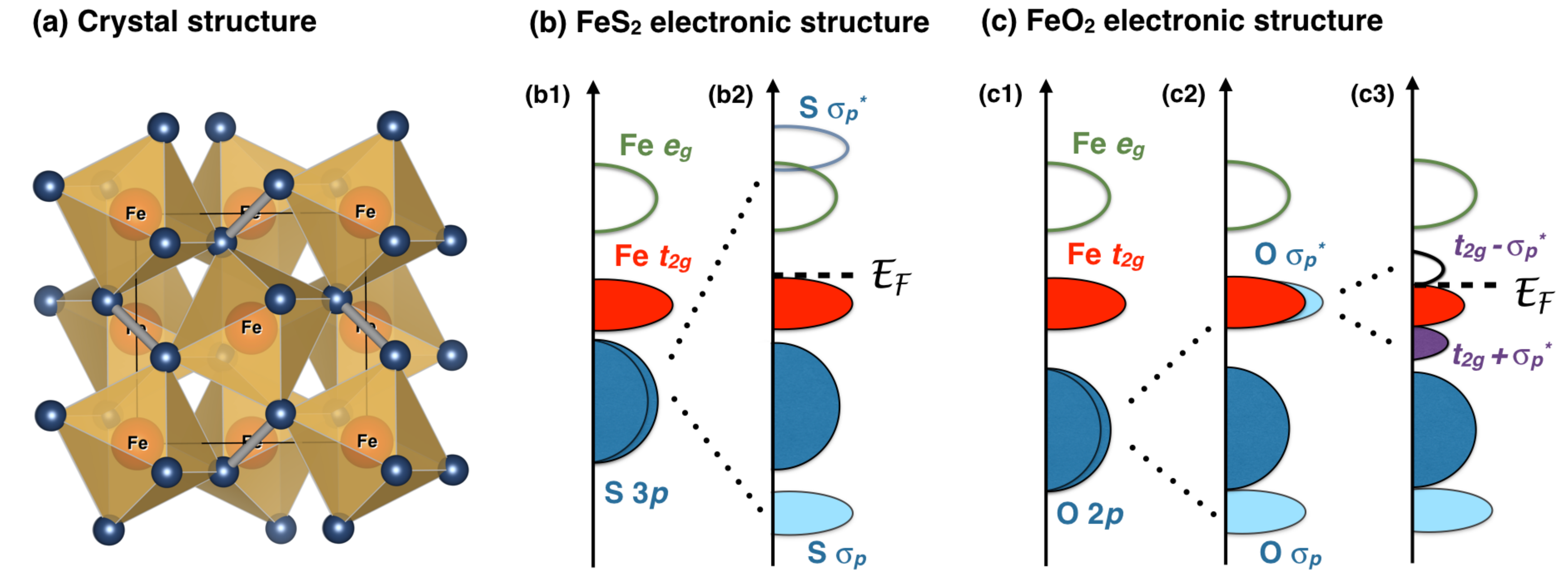}
\caption{(a) Crystal structure of FeO$_2$ and FeS$_2$ can be visualized as a rocksalt structure like FeO with 
O ions replaced by S$_2$ (in FeS$_2$) or O$_2$ (in FeO$_2$) dimers. Fe ions are yellow, while O (or S), forming dimers, ions shown in blue. (b) and (c): Schematic band structure of FeS$_2$ and FeO$_2$.}
 \label{DFT-sketch}
 \end{figure}

Iron peroxide was found to have the same pyrite crystal structure as FeS$_2$\cite{Hu2016}, see Fig.~\ref{DFT-sketch}a, and there is not much difference between oxygen and sulfur from chemical point of view. Thus, it is tempting to consider FeO$_2$ as a complete analogue of FeS$_2$\cite{Jang2017}. Since FeS$_2$ is known to be a diamagnetic insulator with Fe ions adopting 2+ valence state\cite{Burgardt1977,Bullett1976,Ferrer1990},  one might expect that the same is true for FeO$_2$. The first indication that such a picture is oversimplified follows from the recent theoretical study\cite{Jang2017}, where FeO$_2$ was found to be metallic at the pressures where it does exist.

In the present paper we describe electronic and magnetic properties of FeO$_2$. We show that FeO$_2$ is completely different from FeS$_2$,  and so are the physical properties of these compounds. The oxidation state of Fe ion in FeO$_2$ is not 2+, as in FeS$_2$, but close to 3+. This strongly affects magnetic properties of FeO$_2$, since having $3d^5$ electronic configuration,  Fe$^{3+}$ ions may have a magnetic moment. Our comprehensive theoretical calculations using combination of the density functional and dynamical mean-field theories (DFT+DMFT)  demonstrate that there is indeed a highly nontrivial temperature dependence of the magnetic susceptibility in FeO$_2$. We found out that the origin of the difference in magnetic properties between FeO$_2$ and FeS$_2$ and of the metallic character of FeO$_2$ is a much smaller bonding-antibonding splitting for ligand $\sigma$ orbitals in the  peroxide dimer O$_2$ as compared with S$_2$, and a total shift of oxygen $2p$ levels relative to  $3p$ levels of sulfur. This feature of the electronic structure is rather general and important for other dioxides, which can exist in Earth's mantle or in inner parts of exoplanets.

We start with FeS$_2$,  electronic and magnetic properties of which are well understood. As discussed above, one might naively expect to have Fe$^{4+}$ ions with $3d^4$ electronic configuration in FeS$_2$, since usually sulfur has a valence 2-. This would shift Fe $3d$ band very low in energy, below S $3p$, and would result in a self-doping and a metallic conductivity\cite{Sawatzky2016}, which strongly disagrees with the experimental fact that FeS$_2$ is a semiconductor\cite{Bullett1976,Ferrer1990}.

The explanation of this contradiction lies in the specific features of its crystal structure, namely the presence of the S$_2$ dimers. There are sulfurs $3p$ orbitals, which are directed exactly to each other in these dimers. They form such a  strong bond that the antibonding $\sigma^*_p$ orbitals turn out to be higher in energy than the Fe $e_g$ orbitals, see Fig.~\ref{DFT-sketch}(b2). This leads to a formal valency of sulfur ``1-'', (or to (S$_2$)$^{2-}$ dimers), 
and Fe ions become ${2+}$ with the $3d^6$ electronic configuration. Fe ions are in the ligand octahedra in pyrite structure. Strong crystal field splitting between the $t_{2g}$ and $e_g$ bands ($\sim 3.5$ eV in case of FeS$_2$, see Supplemental materials - SM\cite{SM}) counteracts the  Hund's rule and stabilizes the low spin configuration with all six $3d$ electrons occupying $t_{2g}$ sub-shell. This makes FeS$_2$ diamagnetic and insulating\cite{Eyert1998a}.
\begin{table}[b!]
\centering \caption{\label{comparison} Comparison of different physical properties of FeS$_2$ and FeO$_2$, as follows from the DFT and DFT+DMFT calculations.}
\vspace{0.2cm}
\begin{tabular}{lccccc}
\hline
\hline
         &  Fe valence &  Electric properties &  Magnetic properties\\
\hline
FeS$_2$  & 2+          & insulator            &  diamagnetic\\
FeO$_2$  & 3+          & metal                &  paramagnetic\\
\hline
\hline
\end{tabular}
\end{table}

The electronic structure of FeO$_2$ is rather different from a sulfide counterpart. We sketched how this difference appears in Fig.~\ref{DFT-sketch}c (while the results of the actual calculations performed within generalized gradient approximation, GGA,  as well as the details of such calculations are presented in Fig.~S1 in SM\cite{SM}), starting from the hypothetical FeO$_2$ having FCC lattice (like NaCl), where O ions do not form dimers and where there are basically three bands O $p$, Fe $t_{2g}$, and Fe $e_g$, see Fig.~\ref{DFT-sketch}(c1).

\begin{figure}[t!] 
 \centering
\includegraphics[width=0.8\columnwidth]{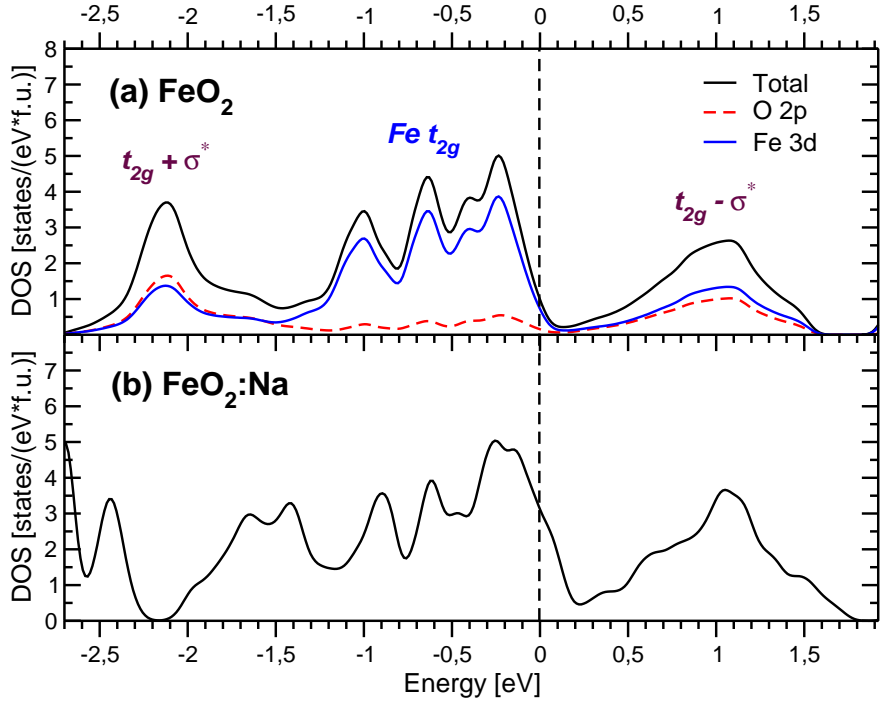}
\caption{Total and partial density of states (DOS) in the nonmagnetic GGA calculations (a) for FeO$_2$ and (b) FeO$_2$ doped by Na (25\%).  Fermi energy is in zero.}
 \label{DFT-bands}
 \end{figure}

First of all, as follows from our GGA calculations, the oxygen $2p$ levels are shifted down relative to the Fe $3d$ states, as compared with the $3p$ levels of sulfur. Besides, as was mentioned above, the presence of the ligand-ligand dimers in real FeO$_2$ results in bonding-antibonding splitting, but since oxygen $2p$ orbitals are much less extended than sulfur $3p$ orbitals, this bonding-antibonding splitting in the O$_2$ dimer is expected to be  
much smaller. As a result the antibonding O $\sigma^*_p$ orbital appears not above $e_g$ (like in FeS$_2$), but exactly in the place, where Fe $t_{2g}$ bands lie, see Fig.~\ref{DFT-sketch}(c2). Then, first of all, part of the Fe $t_{2g}$ electrons would be transferred to oxygens,  shifting Fe valence in the direction of 3+. Second, the hybridization between Fe $3d$ and O $\sigma^*_p$ orbitals again makes  bonding and antibonding combinations, which are labeled as $t_{2g}+\sigma^*_p$ and $t_{2g}-\sigma^*_p$ in Fig.~\ref{DFT-sketch}(c3) respectively. The density of states (DOS) plot in the vicinity of the Fermi energy as obtained in conventional GGA is presented in Fig.~\ref{DFT-bands}(a). These $t_{2g}+\sigma^*_p$ and $t_{2g}-\sigma^*_p$ bands are centered at $-2$ and $1$ eV in  Fig.~\ref{DFT-bands}(a). Note that these bands have nearly the same contributions from Fe $t_{2g}$ and O $2p$ ($\sigma^*_p$) states. Moreover, it is clear that peaks below and above the Fermi level are not bonding and antibonding, since there is no contribution from O $2p$ band below $E_F$. These are nonbonding and antibonding states.

This salient feature of FeO$_2$, that the antibonding $\sigma^*$ orbital falls exactly into the Fe $t_{2g}$ band, determines the main  physical properties of FeO$_2$, which are very different from FeS$_2$, see Tab.~\ref{comparison}. First of all, since there appear additional bands at the Fermi level, while the number of electrons is the same, FeO$_2$ is not a band insulator (as FeS$_2$), but a metal.

There are eight $t_{2g}$ bands, each doubly degenerate with respect to spin, below the Fermi energy for the unit cell consisting of four formula units (f.u.), which are occupied by 4 electrons per each Fe ion (Fig.~S1 in SM\cite{SM}). In addition there are four bonding $t_{2g}+\sigma^*_p$ bands with nearly $50\%$ contribution  of the Fe $t_{2g}$ states (see partial DOS presented in Fig.~\ref{DFT-bands}), which adds approximately one more electron to each Fe ions. As a result Fe ions in FeO$_2$ are nearly $3+$ with $3d^5$ electronic configuration, while in FeS$_2$ they are 2+. sdds

In contrast to Fe$^{2+}$, which is nonmagnetic with $t_{2g}^6$ configuration at large pressure, Fe$^{3+}$ ion even in the low-spin state has a magnetic moment. Moreover, the oxygen $\sigma^*$ states are half-filled in FeO$_2$, and thus they can also contribute to the total  magnetic moment.

Second, the Fermi level appears to be in a very specific position. On one hand it is almost in the pseudogap, so that the Stoner criterion for ferromagnetic (FM) is formally not fulfilled,  and this is the reason why magnetic solutions does not survive in the GGA (we also  checked stability of magnetic solutions at other $q$-vectors, corresponding to AFM-I and AFM-II magnetic structure of FCC lattice of Fe ions\cite{smart-book}). On the other hand it is just on the border line between bands corresponding to localized $t_{2g}$ electrons and antibonding molecular $t_{2p}-\sigma^*_p$ states. This is very important for magnetic properties of stoichiometric and non-stoichiometric FeO$_2$ as we will show latter.

While conventional DFT is exceptionally useful for understanding of the basics of the electronic structure in FeO$_2$, it does not take into account strong Coulomb correlations, which are known to be important for description of the physical properties of many transition metal compounds. We treated correlation effects using the DFT+DMFT method\cite{Anisimov97}. Hubbard $U$ was calculated to be 6 eV, $J_H=0.9$ eV, other details can be found in SM\cite{SM}. 

Correlation effects manifest themselves basically via the renormalization of the GGA DOS near the Fermi level,  $m^*/m \sim$ 1.2-1.6 (depending on the orbital), resulting spectra functions are shown in Fig.~S2 of SM.  FeO$_2$ is a bad metal for experimental pressure of 76 GPa.  There are 4.8 electrons in the $t_{2g}$ shell, which certifies that Fe is 3+ in FeO$_2$. The local magnetic moment $\langle \sqrt {m_z^2} \rangle$ was found be 1.5 $\mu_B$. The contribution from the $t_{2g}$ orbitals to the total local magnetic moment, $\langle m_z^2 \rangle_{t_{2g}} = 1.08 \mu_B^2$, exactly corresponds to the low spin state of $3d^5$ configuration. There is, however, also an additional contribution, $\langle m_z^2 \rangle_{e_{g}} = 1.04 \mu_B^2$, due to a partial polarization of the ligand electrons residing $e_g$ shell of transition metal (see detailed discussion in Supplemental materials). In spite of the fact that there are magnetic moments on Fe ions, they do not order, so that FeO$_2$ stays paramagnetic down to 190 K  (we checked FM and AFM-I). Even lower temperatures can be reached in our calculations by using a truncated Hamiltonian, which includes only Fe $t_{2g}$ and O $2p$ states (this choice of impurity orbitals gives the same spectral functions in vicinity of the Fermi level and very similar $\chi(T)$ as full $3d$ Hamiltonian). 
 In this case we were able to go down to 60 K,  and again FeO$_2$ does not order in our calculations even at these temperatures. This may seem somewhat surprising since having a rather large bandwidth (and hence hopping parameters) one might expect large superexchange interaction between Fe ions, if spins would have been localized.

\begin{figure}[t!] 
\centering
\includegraphics[width=1\columnwidth]{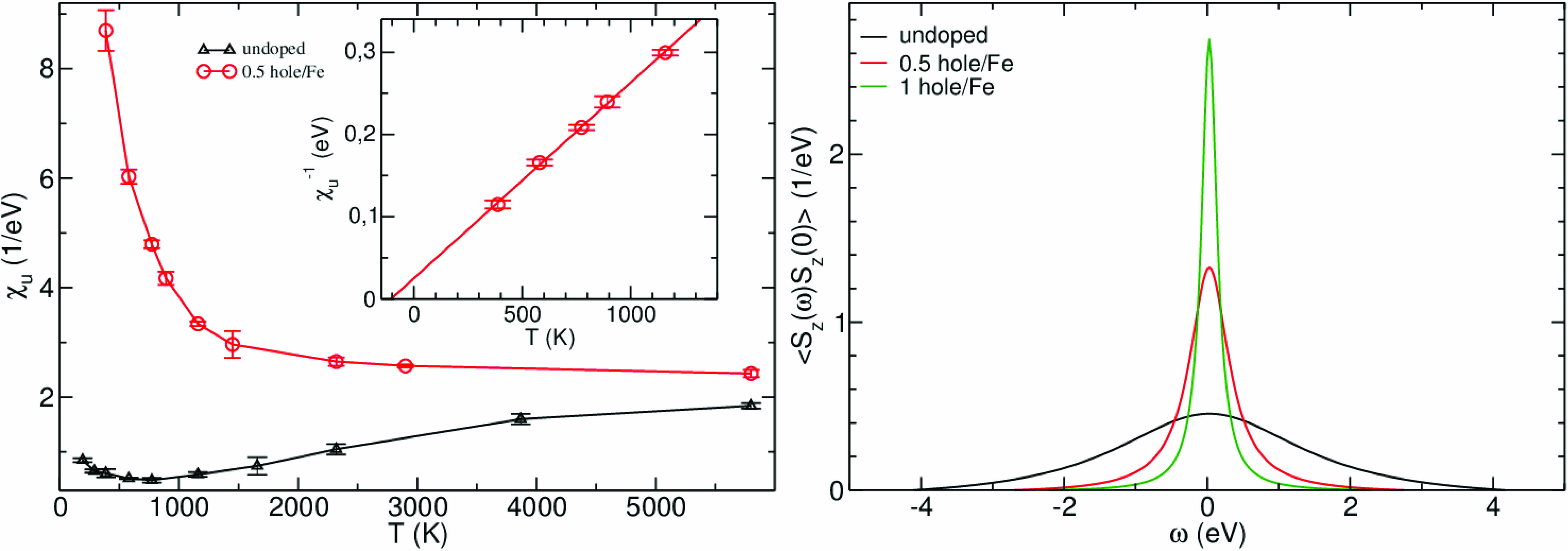}
\caption{Results of the DFT+DMFT calculations. Left panel:
uniform magnetic susceptibility for pure FeO$_2$ and $\frac{1}{2}$ hole/Fe doping (red circles). Inset shows magnetic susceptibility for 0.5 hole per Fe atom as a function of 1/T. Right panel shows local magnetic susceptibility as a function of frequency for different doping.
}
 \label{DMFT-chi}
 \end{figure}

In order to estimate the degree of the spin localization we calculated the analytical continuation on real frequency of the spin-spin correlator $\langle S_z(i\omega)  S_z(o) \rangle =  \int_0^{1/k_BT}d\tau\langle S^z(\tau) S^z(o) \rangle e^{i\omega_n\tau}$, where $\tau$ is an imaginary time, see right panel in Fig.~\ref{DMFT-chi}~\cite{Katanin2010,Georges-96}. 
The width of this correlator is inversely proportional to the lifetime of a magnetic moment. For example in a pure metallic iron, where $t_{2g}-e_g$ crystal field splitting is small, iron ion is in a high-spin state. The magnetic moment can be localized,  with the full width at half maximum (FWHM) of about 0.2~eV for the less localized $\gamma$-Fe and 0.1~eV for the more localized $\alpha$-Fe~\cite{Katanin2010,Igoshev2013}. From the inset of  Fig.~\ref{DMFT-chi} one may see that in FeO$_2$ FWHM of the spin-spin correlator is $\sim 3$ eV, which demonstrates that the magnetic moments can hardly be considered as localized.

In DMFT one can calculate the uniform magnetic susceptibility $\chi_u(T)$ as a response to an external magnetic field, which is introduced via Zeeman splitting $\delta E$ in the Hamiltonian:
\begin{eqnarray}
\chi_u(T) = \frac {\delta m}{\delta H} = \frac {n^{\uparrow} - n^{\downarrow}}{\delta E} \mu_B^2.
\end{eqnarray}
Here $m$ is the magnetization, $n^{\uparrow}$ and $n^{\downarrow}$ are total occupations for spin up and down. This direct calculation of the uniform magnetic susceptibility, $\chi_u(T)$ shows that it has a nontrivial temperature dependence. Namely, with increasing temperature  $\chi_u$ first decreases (for $T<T^*=$750 K), and then starts to increase almost linearly above $T^*$, which resembles the behaviour of the pnictides\cite{Skornyakov2011}. Detailed analysis of these data\cite{SM} shows that such an unusual for 3D system behavior is due to a specific position of the Fermi level in between the localized $t_{2g}$ and antibonding $t_{2g} - \sigma_p^*$ states. At low temperature the particle-hole excitations occur within the localized Fe $t_{2g}$ states and $\chi_u(T)$ goes down with temperature, resembling the Curie-Weiss law. Increasing temperature further ($T>T^*$ K),  we start to excite molecular-like $t_{2g} - \sigma_p^*$ states, which leads to a completely different temperature dependence.

This means that the electron and hole doping, which is likely to occur in Earth's mantle, would result in a very different temperature dependences of magnetic susceptibility, since we shift the Fermi level to the peaks corresponding to {\it qualitatively} different states (localized and molecular-like). There are many different elements besides Fe (5.8\%) and O (44.8\%) in the Earth's mantle, and one may expect that Mg ($\sim$ 22.8\%), Si ($\sim$21.5 \%), Ca ($\sim$2.3 \%) or Na (0.3\%)\cite{jackson-book} may dope FeO$_2$ and change its properties dramatically. Indeed, the Fermi level in stoichiometric FeO$_2$ is on the steep slope of a large peak in DOS, and changing its position we strongly affect both magnetic and electronic properties.

The electron doping will shift the Fermi level to antibonding molecular-like $t_{2g} - \sigma_p^*$ states, which is unlikely to provide a large magnetic response in the simplest rigid-band shift model. Moreover, by doing this we transform Fe ion into the nonmagnetic low-spin $3d^6$ configuration, corresponding to the 2+ oxidation state, so that only a small electron doping can increase magnetic moment. In addition the electron doping is rather unfavourable from structural point of view: the population of the strongly antibonding $t_{2g} - \sigma_p^*$ orbital would significantly weaken O$_2$ dimers, existing in the pyrite structure. Thus, at first sight the hole doping is expected to be much more effective for making FeO$_2$ magnetic: the Fermi level would then be  shifted to the large peak corresponding to localized Fe $t_{2g}$ electrons.

\begin{figure}[t!] 
 \centering
\includegraphics[width=1\columnwidth]{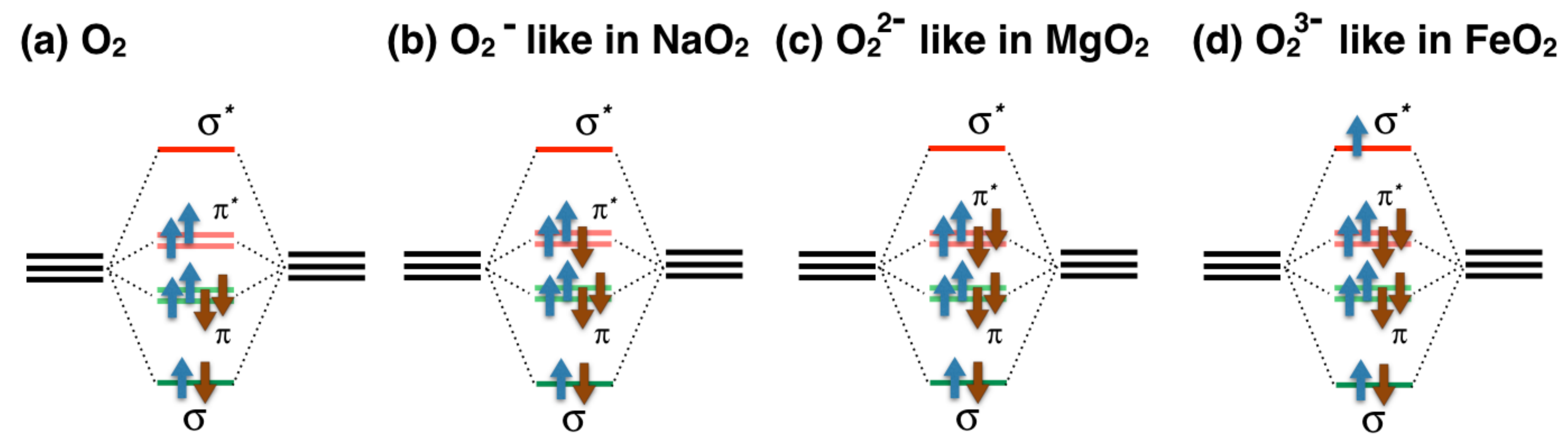}
\caption{Occupation of oxygen $2p$ orbitals in different compounds with O$_2$ dimers. The system gains (loses) energy by occupation of green bonding (red antibonding) bonds.}
 \label{O-sketch}
 \end{figure}

We checked different types of hole and electron dopings by the GGA calculations (for ferromagnetic order) performing full structural optimization, starting from the pyrite structure and substituting 25\% of Fe by different ions such ions as Mg, Si, and Na.  Mg doping formally changes valence of the peroxide O$_2$ group from 3- in FeO$_2$ to 2- in MgO$_2$ (see Fig.~\ref{O-sketch} (c) and (d)), but it has no influence either on band structure or on magnetic properties of the system: unoccupied  $\sigma^*$ band corresponding to the Mg(O$_2$)$^{2-}$ unit appears just above the Fermi level and does not provide any holes to the Fe ions. In NaO$_2$ superoxide
the O$_2$ ``molecule'' is 1-, see Fig.~\ref{O-sketch}(b) and also Ref.\cite{Solovyev2014}, and hence by Na we depopulate O $\pi^*$ bond, which will be immediately refilled by the Fe $t_{2g}$ electrons. This leads to the shift of the Fermi level downwards, see Fig.~\ref{DFT-bands}b, and results in the magnetic instability. In the GGA calculations the magnetic moments on Fe ions were found to be $\sim$0.4$\mu_B$. Si doping keeps FeO$_2$:Si nonmagnetic, but only in unrelaxed crystal structure. After lattice optimization there appears two very different O$_2$ dimers, which help to form magnetic moment $\sim$0.4$\mu_B$ even in the case of the light electron doping. But the most effective are Fe vacancies (25\%), which give magnetic ground state in the GGA calculations with magnetic moments $\sim$0.6$\mu_B$.

Thus, we see that there are plenty of possibilities for FeO$_2$ to be magnetic due to different types of doping or because of non-stoichiometry. It is hard to expect, however, that FeO$_2$ would order magnetically in the Earth's mantle, because of very high temperatures, $\sim$1000-2000 K, but even in a paramagnetic state it may still provide local magnetic moments. The direct DFT+DMFT calculations within the rigid-band shift approximation (as one can see from Fig.~\ref{DFT-bands}b, the band structure does not change dramatically with doping) show the drastic increase of the uniform magnetic susceptibility with hole doping, see Fig.~\ref{DMFT-chi}. it is now Curie-Weiss like in a wide temperature range and  the spin-spin correlation function demonstrates an increase of the local magnetic moments lifetime (i.e. decrease of the width of the correlator, see inset in Fig.~\ref{DMFT-chi}) with doping. 
 
 In addition to a possible importance of our findings for geoscience, FeO$_2$ represents an exceptional interest also for physics and solid state chemistry, since it lies on the borderline between the stable dioxides of transition metals, such as TiO$_2$ VO$_2$, CrO$_2$ etc.,  and equally stable oxides and sulfides having pyrite structure, such as NaO$_2$, KO$_2$, FeS$_2$ etc.  FeO$_2$ may thus be considered as a ``bridge'' between dioxides and peroxides/disulfides, and it displays properties of both.

There is a well known concept in physics, introduced by Zaanen, Sawatzky and Allen\cite{ZSA}, that going along a row in the periodic table from the left to the right, or increasing valence of a metal in a transition metal oxide, we go over from the Mott insulator, where the band gap is defined by Hubbard $U$, to a charge-transfer regime, where it is given by the charge transfer (CT) from a ligand to a metal, $\Delta_{CT}>0$, and finally to the state, where  $\Delta_{CT}$ becomes negative with ligands donating some of their electrons to a metal (so called self-doping) \cite{ZSA,khomskii2014transition}.

In peroxides the situation with the CT energy is ``inverted'' from the beginning: as we have seen in FeS$_2$ part of electrons are transferred from sulfur to Fe. CoS$_2$, NiS$_2$, MgO$_2$, KO$_2$ and many other materials are just the same: ligand $\sigma^*$ and sometimes even $\pi^*$ orbitals donate (see Fig.~\ref{O-sketch}) at least one electron for a metal, i.e. oxygen is 1- or even 1/2-. With FeO$_2$ one returns to normal transition metal oxides, where oxygen's valency is 2-, but there is still one step to make since O is 1.5- in FeO$_2$. Thus, we see that FeO$_2$  indeed lies ``in between'' oxides and peroxides/disulfides, which makes it an especially interesting material from physical point of view.

A simple qualitative difference between normal (di)oxides  and peroxides  is the following: On one hand, when the main ``structural unit'' in a system is a single O ion, like in dioxides of the type of TiO$_2$, VO$_2$, its ``natural'' state is O$^{2-}$, and counting from that, we see that e.g.  in FeO$_2$ the CT energy would be negative, $\Delta_{CT}<0$, i.e. the electrons would be transferred from O$^{2-}$ to Fe (as it happens already in CrO$_2$\cite{Korotin1998}). But in peroxides, as well as, e.g., in FeS$_2$, the natural ``structural unit'' is the O$_2$ or S$_2$ dimer. Such dimer can be in different charge states:  neutral O$_2$  molecule, Fig.~\ref{O-sketch}a;  (O$_2$)$^-$ molecular ion (say in NaO$_2$, KO$_2$), Fig.~\ref{O-sketch}b; or (O$_2$)$^{2-}$ ion as in MgO$_2$, Fig.~\ref{O-sketch}c.

The more electrons we put on such a dimer, the more we fill antibonding states, which gradually destabilizes the very O$_2$ dimers. But till (O$_2$)$^{2-}$ it is still reasonably harmless, we fill ``weakly'' antibonding states ($\pi^*$), see Fig.~\ref{O-sketch}.  But as soon as one starts to occupy the upper $\sigma^*$ states, the very dimers start to become more and more destabilised, which we indeed see in FeO$_2$: the O-O distance in (O$_2$)$^{3-}$ dimers is 1.89 \AA\cite{Hu2016}~- much larger than  1.49 \AA~for (O$_2$)$^{2-}$ dimer in MgO$_2$\cite{Vannerberg1959} or 1.32 \AA~ for (O$_2$)$^-$ in NaO$_2$\cite{Ziegler1976}.  Already MgO$_2$, having 4 electrons on antibonding $\pi^*$ orbitals, see Fig.~\ref{O-sketch} (c), readily decomposes at zero pressure\cite{Wriedt1987}. In FeO$_2$ we lose even more energy occupying antibonding $\sigma^*$ orbital, see Fig.~\ref{O-sketch} (d). This makes FeO$_2$ even less stable in the pyrite structure, than MgO$_2$, so that it can be stabilised only at a very high pressure.

Summarising, we see that the recently discovered pyrite-like FeO$_2$\cite{Hu2016} is even more exotic than it was initially thought. Unexpected valence states, nontrivial magnetic properties, stabilization of local magnetic moments by non-stoichiometry or doping by such abundant constituents of Earth's mantle such as Si (and Na)
and finally its special place between (di)oxides  and peroxides make FeO$_2$ extremely interesting not only for geoscience, but also for the condensed matter physics and solid state chemistry.


\section*{Acknowledgements (not compulsory)}
We thank V. Irkhin and I. Mazin for various stimulating discussions. The study of electronic and magnetic properties of FeO$_2$ performed by S.V.S. was supported by the Russian Science Foundation (project no. 17-12-01207), while work of D.I.Kh. was supported by the Deutsch Forschungsgemeinschaft through the CRC 1238 program. 

\section*{Data availability statement}
No datasets were generated or analysed during the current study.

\section*{Author contributions statement}
S.V.S. conceived the model describing electronic structures of FeO$_2$ and FeS$_2$. S.V.S. and D.I.Kh. discussed and analysed implications of features of the electronic structure for physical properties of FeO$_2$. The DFT calculations were performed by S.V.S. and A.O.Sh. The DFT+DMFT calculations were carried out by A.O.Sh.; A.I.P., A.O.Sh., S.V.S., and S.L.S. analysed magnetic properties of FeO$_2$. S.V.S. and D.I.Kh. wrote the manuscript with help of all co-authors.

\section*{Additional information}
The authors declare no competing financial interests.

\end{document}